\newcommand{\ba}{BaCu$_2$Si$_2$O$_7$}

\documentstyle[prl,aps,twocolumn]{revtex}
\begin{document}
\input{psfig.sty}
\draft

\twocolumn[\hsize\textwidth\columnwidth\hsize\csname
@twocolumnfalse\endcsname

\title
{Interacting quantum spin chains.\\ Invited paper for the
International Conference on Neutron Scattering ICNS-2001. }

\author{A. Zheludev}

\address{Physics Department, Brookhaven National Laboratory, Upton, NY
11973-5000, USA. \\ New address: Physics Department, Oak  Ridge
National Laboratory, Oak Ridge, TN 37831-6393, USA. }

\date{\today}
\maketitle
\begin{abstract}
A brief review of recent advances in neutron scattering studies of
low-dimensional quantum magnets is followed by a particular
example. The  separation of single-particle and continuum states
in the weakly-coupled $S=1/2$ chains system \ba\ is described in
some detail.\\
\end{abstract}

]

For the last two decades low-dimensional quantum magnets have been
the subject if intensive neutron scattering studies. One of the
main reasons for this steady interest is that low dimensional
systems are {\it simple} models of magnetism, that demonstrate a
broad spectrum of {\it complex} quantum-mechanical phenomena. In
many cases quantum magnets are desribed by simple Hamiltonians
with few parameters. Theoretical and numerical studies of these
models can be directly compared to experiment at the quantitative
level, often yielding remarkable agreement, and provide guidance
in the data analysis.  Neutron scattering techniques are
particularly well suited for studying real low-dimensional
magnets. Indeed, they provide {\it direct} measurements of the
spin correlation function $S(\bbox{q}, \omega)$, that carries
significant physical information and is the ultimate result of
most theoretical calculations. Moreover, in most known
low-dimensional magnets the energy and length scales of magnetic
interactions perfectly match those probed by thermal or cold
neutrons. It will not be an overstatement to say that the
development of the entire field of low-dimensional magnetism has
been driven by neutron experiments more than by any other
experimental technique.

Two decades of research and huge amounts of beam time yielded a
fairly complete understanding of the most basic one-dimensional
models. To mention only a few milestones, we have to recall the
study of local excitations in dimer systems,\cite{dimers} the
discovery of the famous Haldane gap\cite{Haldanegap} and the
observation of continuum excitations in $S=1/2$ Heisenberg
antiferromagnets (AFs) \cite{continuum,Dender96}. A number of
remarkable discoveries were made only recently. Among these are
studies of multi-magnon excitations \cite{multi-magnon},
observation of field-induced incommensurability in $S=1/2$ systems
\cite{cube}, the spin-Peierls compound CuGeO$_3$ \cite{cugeo3},
continuum states\cite{Haldanecontinuum} and field-induced
ordering\cite{Haldanefield} in Haldane-gap antiferromagnets, and
the effect of staggered fields on quantum spin chains
\cite{Haldanestaggered}. These new studies were enabled by the
discovery of new model materials, development of new experimental
techniques and the perfection of data analysis procedures.

Today, the general trend in low-dimensional magnetism is to
capitalize on the accumulated knowledge of the basics and move on
to more complex problems. Among the new and rapidly progressing
directions of research are effects of randomness and doping in
quantum spin chains \cite{YBANOdoping,PBNI,CUGEOdoping}, the
interplay between charge and spin degrees of freedom
\cite{charge}, new physics in highly frustrated quantum
antiferromagnets \cite{frustrated}, and the crossover regime from
``quantum'' to ``classical'' magnetism. In the talk we will
attempt to cover as many of these new studies as possible. To keep
the present paper at least marginally readable however, below we
shall concentrate on just one example, namely the dimensional
crossover in weakly-interacting $S=1/2$ Heisenberg spin chains.

At the heart of the matter is a very old controversy. As far back
as 1931 H. Bethe {\it exactly} solved the ground state of the
one-dimensional (1D) $S=1/2$ quantum Heisenberg antiferromagnet
\cite{Bethe31}. The main result was that even at $T=0$ there is no
long-range order in the system, and no Bragg peaks should be
visible in a neutron diffraction experiment. A year later, L.
N\'{e}el proposed the famous two-sublattice model of
antiferromagnetism \cite{}, characterized by staggered long-range
magnetic order, that produces new magnetic Bragg peaks in the
diffraction pattern. In 1933 L. Landau published yet another paper
on the subject, critisizing the 2-sublattice model based on the
fact that it is not even an eigenstate of the Heisenberg
Hamiltonian, and therefore can not {\it possibly} be the ground
state \cite{}. Now we of course know that for a vast majority of
2- and 3D materials, the ground state does indeed look remarkably
{\it like} the Neel state. Landau's arguments are {\it also}
correct, and quantum fluctuations are relevant. In 2 and 3
dimensions they usually result in minor corrections. The lower the
effective dimensionality, the more these fluctuations are
important, and in the purely 1D case they are capable of
destroying long-range order altogether. It is now well understood
that {\it weakly coupled} $S=1/2$ Heisenberg chains are {\it
weakly ordered}: the Neel temperature $T_N$ scales roughly as the
strength of inter-chain coupling $J'$, while the sublattice
saturation moment at $T\rightarrow 0$ behaves as $J'/J$, $J$ being
the in-chain exchange constant. Both quantities vanish as
$J'\rightarrow 0$.  It is important to note that long-range
ordering occurs for arbitrary small $J'$. For example, correlated
glassy freezing with an ordered moment of only 0.03~$\mu_B$ have
recent been detected in the {\it extremely} one-dimensional
material SrCuO$_2$ with $J'/J\approx 7 \times 10^{-4}$
\cite{Zaliznyak99}.

\begin{figure}
 \psfig{file=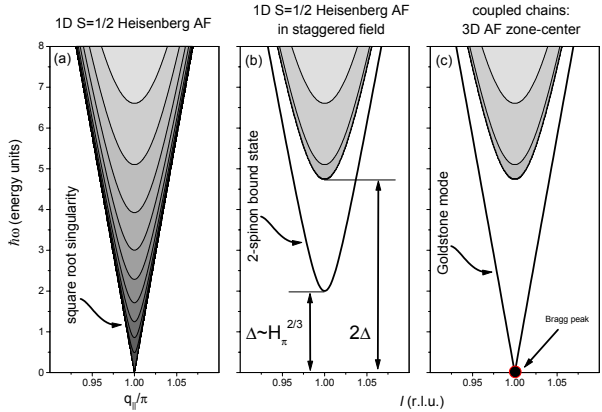,width=3.3in,angle=-90}
\caption{
 Transverse dynamic structure factor of the 1D $S=1/2$ Heisenberg AF
 (a)contains only continuum excitations with a singularity on the
 lower bound. An external staggered field (b) induces a gap
 $\Delta$ in the spectrum. The singularity separates from the
 lower bound of the continuum giving birth to single-particle
 excitations. This picture is also observed in coupled chains at
 the transverse zone-boundary. In the latter case the single-particle states take
 the role of Goldstone modes (spin waves) and their energy
 goes to zero at the  3D AF zone-center (c).}
 \label{cartoon}
\end{figure}

The most interseting question is what happens to the excitation
spectrum of a single $S=1/2$ antiferromagnetic quantum spin chain
when inter-chain coupling is ``switched on''. Let us first
consider the extreme cases. In the 3D limit, when $J'\approx J$ we
are dealing with a ground state that is very similar to the Neel
state. The excitation spectrum is then dominated by
single-particle states that correspond to a {\it presession} of
the ordered moment around its equilibrium direction. These
particles, known as {\it spin waves}, carry a total spin of unity
and and a spin projection $S_z=\pm 1$ onto the direction of
staggered moment. In the early days it was believed that the other
limiting case of a purely 1D AF the excitation spectrum is
described by a similar single-particle picture, albeit with
strongly renormalized spin wave velocity and bandwidth \cite{DCP}.
It was later realized that spin dynamics in the 1D case is, in
fact, {\it qualitatively} different. Since long-range order is
absent, so are the precession modes. The spectrum contains {\it
no} single-particle excitations and is instead a continuum of
states \cite{Fadeev81,Muller,Haldane93,Karbach97}. An experimental
confirmation of this phenomenon was obtained in elegant neutron
scattering experiments on KCuF$_3$ \cite{continuum} and copper
benzoate\cite{Dender96}. Modern theories describe these continuum
states as composed of pairs of exotic $S=1/2$ excitations called
{\it spinons}. Unlike magnons, which are bosons and can be
directly observed in an inelastic neutron experiment, spinons are
fermions and are created or destroyed only in pairs, much like
domain walls in an Ising magnet. The two-spinon continuum is
3-fold degenerate with pairs of spinons having a total spin of
unity and a projections on any given axis $S_z=0,\pm 1$. Note that
while there are only two polarizations for spin waves, spinon
pairs come in three different polarization flavors.

If the spin dynamics in the two limiting cases is {\it
qualitatively} different, what happens in quasi-1D systems with
$0<J'\ll J$? The presence of long-range order should produce
order-parameter excitations, i.e., spin waves. But how exactly are
these single-particle states spawned from the continuum of
inelastic scattering that dominates in the 1D systemmodel? A
simple physical picture is provided my the chain-mean field (MF)
theory \cite{Scalapino75}. In the ordered state each spin chain is
subject to an effective staggered exchange field generated by
neighboring chains. A staggered field $H_\pi$ induces a liner
attractive potiential between spinons. As a result, the
lowest-energy excitations are spinon {\it bound states}, often
referred to as ``magnons''  \cite{Schulz96,Essler97}. This is
illustrated in Fig.~\ref{cartoon}(b). The square root singularity
on the lower bound of the 2-spinon continuum in the isolated
chains [Fig.~\ref{cartoon}(a)] ``separates'' and becomes a sharp
magnon which is a $\delta$-function in energy at any given wave
vector [Fig.~\ref{cartoon}(b), solid line]. The magnons acquire a
{\it gap} $\Delta$ (also referred to as {\it mass}), that scales
as $H_\pi^{2/3}$. Since there are three possible spin states for a
pair of spinons, there are {\it three} magnon branches. Two of
these are polarized perpendicular to $H_\pi$ and the induced
staggered moment, and correspond to conventional precession modes
(spin waves). Including inter-cahin interactions within the Random
Phase Approximation (RPA) gives these excitations a dispersion
perpendicular to the chains. Their energy goes to zero at the 3D
zone-center, i.e., at the location of magnetic Bragg peaks in the
ordered system [Fig.~\ref{cartoon}(c), solid line]. The gap
$\Delta$ can still be observed at the transverse zone-boundary,
where the behavior of an isolated chain in a staggered field is
exactly recovered [Fig.~\ref{cartoon}(b)]. What remains of the
2-spinon continuum in the 1D system is now seen as a {\it
2-magnon}, rather than {\it 2-spinon} continuum. Indeed, the
attractive potential between spinons is a {\it confining} one, and
two spinons are permanently bound into magnons, just like two
quarks can be confined in a meson. The continuum therefore has a
gap of to $2\Delta$, {\it i. e.}, twice the characteristic magnon
gap.

An experimental observation of such rich and unique behavior, the
{\it separation} of single-particle and continuum states, is a
formidable challenge to neutron scattering. On the one hand, a
strongly 1D system with $J'\gg J$ is desirable to maximize the
fraction of the spectral weight contained in the continuum, a
feature notoriously difficult to observe. On the other hand, $J'$
should be large enough to yield a measurable gap $\Delta$
(preferably, a few meV). Finally, $J$ should be {\it small} enough
to allow measurements with a wave vector resolution better than
$\Delta/v$, where $v=\pi/2J$ is the spin wave velocity. The two
latter conditions are absolutely essential to resolving the
magnons at energy $\Delta$ from the lower bound of the continuum
at $2\Delta$. The first model system that met these conflicting
requirements was KCuF$_3$, a material with $J=??$, $T_{\rm N}=??$
and a saturation moment of $m_0\approx 0.5~\mu_B$. In this
compound the spin waves and continuum excitations could be
observed simultaneously \cite{Tennant95}.

\begin{figure}
 \psfig{file=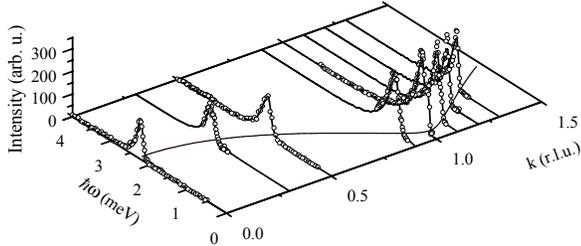,width=3.3in,angle=0}
\caption{A series of constant-$Q$ scans measured in \ba\ at
$T=1.5$~K for different momentum transfers perpendicular to the
chain axis. Lines represent a semi-global fit to the data as
described in the
 text. The solid lines in the basal plane show the spin wave dispersion relation in this
 reciprocal-space direction. The data are from Ref.~\protect\cite{Zheludev00Ba}} \label{exdata1}
\end{figure}

Below we shall make the experimental case for separation of
single-particle and continuum states using another model quasi-1D
material, namely \ba. In this compound $J=24$~meV, $T_N=9$~K and
$m_0=0.15~\mu_B$ \cite{Tsukada99,Kenzelmann01}, {\it i. e}., \ba\
is more 1-dimensional than KCuF$_3$. The $S=1/2$ AF chains run
along the $c$ axis of the orthorhombic crystal structure. The 1D
AF zone-center $q_\|=\pi$ is the $(h,k,1)$ reciprocal-space plane,
and the magnetic Bragg peak, characteristic of  3D long-range
ordering is located at (0,1,1). Despite the small saturation
moment in \ba, its low-energy excitation spectrum (up to about
5~meV energy transfer) is entirely dominated by sharp
single-particle spin-wave like excitations
\cite{Zheludev00Ba,Zheludev01Ba}. Very high resolution
measurements performed using the IN14 cold-neutron spectrometer at
ILL failed to detect any intrinsic excitation widths.
Fig.~\ref{exdata1} shows a series of constant-$q$ scans that
measure the dispersion of these modes at the 1D AF zone-center in
the direction perpendicular to the chain axis. The solid lines in
Fig.~\ref{exdata1} are a {\it global} fit to the data based on a
single-mode cross section for a classical antiferromagnetic spin
wave, convoluted with the spectrometer resolution function
\cite{Zheludev01Ba}. Measurements of the spin wave dispersion
along different reciprocal-space directions led to a fairly
complete picture of inter-chain interaction \cite{Kenzelman01}.
The effective MF inter-chain coupling constant was found to be
$J'=0.4$~meV. The ``magic point'' where inter-chain interactions
cancel out at the RPA level is located at $(0.5,0.5,1)$. The
energy of the spin wave at this wave vector is to be interpreted
as the gap $\Delta$ induced in each individual chain by their
interactions with neighboring chains. Experimentally, for \ba,
$\Delta=2.5$~meV.
\begin{figure}
 \psfig{file=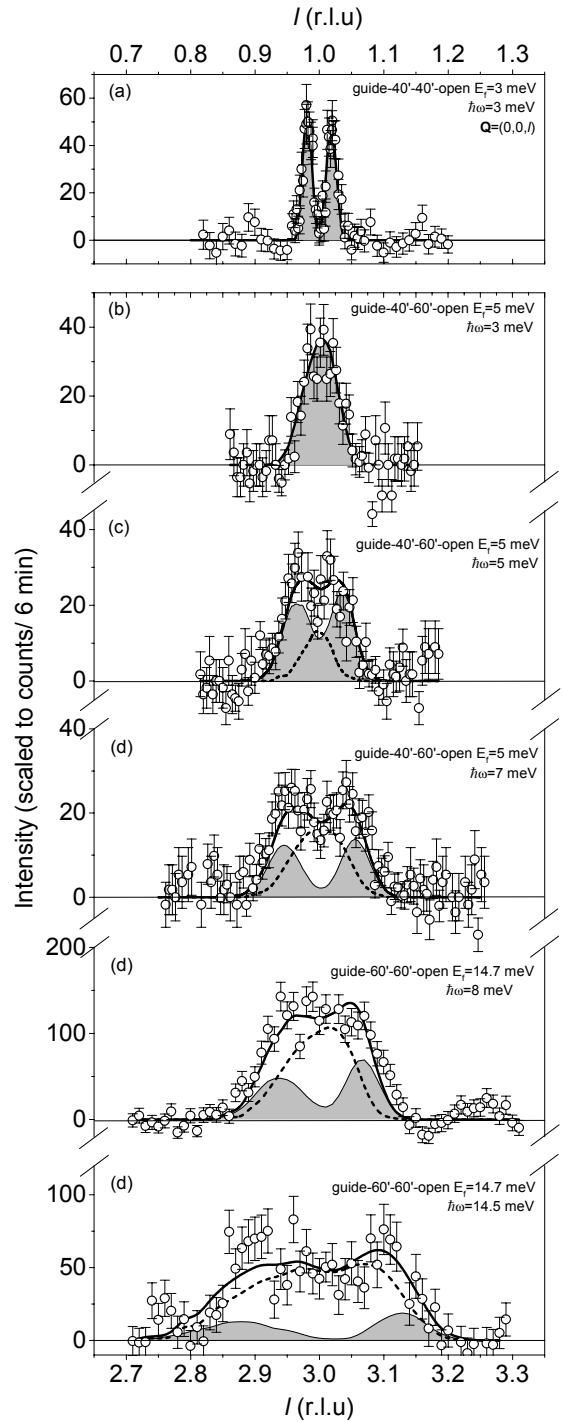,width=3.3in,angle=0}
\caption{
 A series of constant-$E$ scans along the spin chains in \ba.
 Heavy solid
 lines represent a global fit to the data as described in the
 text. Shaded areas are contributions of single-particle
 excitations. Dashed lines show the continuum
 portion. Arrows indicate the slight dip in the observed intensity that
 corresponds to the continuum energy gap $\Delta_c$. The data are from Ref.~\protect\cite{Zheludev01Ba}}
\label{exdata2}
\end{figure}The observed low-energy spectrum is totally
consistent with theoretical predictions: at energies below
$2\Delta$ transverse spin fluctuations in a weakly-couple chains
system behaves exactly as those in a {\it classical}
antiferromagnet.

\begin{figure}
 \psfig{file=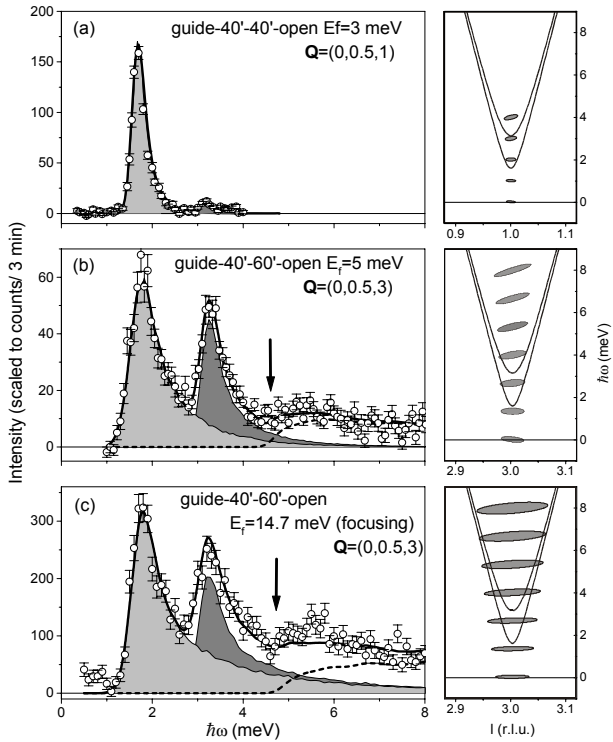,width=3.3in,angle=0}
\caption{
 Left: Typical constant-$Q$ scans collected in \ba\ at the 1D AF
 zone-center using different spectrometer configurations. Lines and shaded areas are as in
 Fig.~\protect\ref{exdata2}. Right:
 Evolution of the calculated FWHM resolution ellipsoids in the
 course of the corresponding scans, plotted in projection onto the $(l,\hbar \omega)$ plane.
 Solid lines represent the spin wave dispersion relation. The data are from Ref.~\protect\cite{Zheludev01Ba}}
 \label{exdata3}
\end{figure}

The quantum-mechanical nature of the spin chains in \ba\ becomes
apparent on shorter time scales (larger energy transfers). Figure
\ref{exdata2} shows a series of constant-energy scans accross the
1D AF zone-center. At $\hbar \omega=3$~meV using the
highest-resolution setup [Fig.~\ref{exdata2}(a)] one clerly sees
two well-resolved peaks that represent the low-energy
single-particle excitations. A fit of the classical spin wave
cross section to the data is shown by the shaded area. The two
spin wave peaks can not be resolved at $\hbar \omega=3$~meV using
a setup with coarser resolution [Fig.~\ref{exdata2}(b)]. However,
at higher energies, [Fig.~\ref{exdata2}(c-f)] even the
coarse-resolution configuration should have been capable of
resolving two separate peaks if the single-particle picture still
held (shaded areas). In contrast, the measured scans do not
contain two separate peaks, but instead show a single broad
feature. Moreover, the spin waves, for which intensity scales as
$1/\omega$, are expected to  account for only a very small
fraction of the total spectral weight at high energy transfers
[Fig.~\ref{exdata2}(e,f)]. The remaining scattering is to be
attributed to the excitation continuum that sets in at about 5~meV
energy transfer and becomes progressively more dominant at high
energies. The bulk of the data collected in different experimental
configurations was analyzed in a global fit using a cross section
that contained both a single-mode and a continuum part. The cross
section for the continuum was chosen to match the theoretical
result of Ref.~\cite{Essler97}. The continuum was assumed to have
a gap of $\Delta_c=2\Delta=5$~meV, i.e., exactly twice the spin
wave gap at the ``magic'' reciprocal-space point. In
Fig.~\ref{exdata2} the result of this global fit is shown in a
solid line, and the continuum contribution is represented by the
dashed line.

The fact that the continuum starts above a well-defined gap energy
$\Delta_c$, can be clearly seen in the wide-range constant-$q$
scans shown in Fig.~\ref{exdata3}. At this wave vector there are
to spin wave peaks due to a non-trivial 3D arrangement of magnetic
ions in \ba (shaded areas). At high energies there is additional
broad scattering not accounted for by the single-particle picture.
The onset of the continuum is signaled by an intensity dip at
around 5~meV (arrows). As in Fig.~\ref{exdata2}, the solid lines
in Fig.~\ref{exdata3} represent the global fit, and the dashed
line is the continuum part of the cross section. If $\Delta_c$ is
treated as an adjustable parameter in the fit, the refined value
is $\Delta_c=4.8(2)$~meV, which is within the error bar of the
theoretical value $\Delta_c=2\Delta$~meV.

The continuum gap being {\it twice} the spin wave gap is a
non-trivial result. All the data discussed above were collected
with scattering vectors almost parallel to the chain-axis. The
ordered moment in \ba\ is parallel to the chains as well, so the
intrinsic polarization dependence of the neutron scattering cross
section ensures that all scans represent {\it
transverse}-polarized spin fluctuations. In conventional SWT the
lowest-energy transverse continuum excitations are {\it
three}-magnon states, since the magnons themselves are
transverse-polarized. In the SWT, the transverse continuum thus
has a pseudogap of $3\Delta$. A rigorous SWT calculation for \ba\
gives $\Delta_c=7.5$~meV \cite{Zheludev00Ba,Zheludev01Ba}. How is
it possible that we are seeing continuum scattering at $2\Delta$?
The answer given by the quantum chain-MF model is that since there
are three possible polarizations for pairs of spinons $S_z=0,\pm
1$ (see above), there is a {\it third} bound state (magnon) that
is polarized parallel to the direction of ordered moment. In a
recent elegant study this {\it longitudinal mode} has been
directly observed in KCuF$_3$ using unpolarized \cite{Lake00} and
polarized \cite{Nagler01} neutrons. The longitudinal magnon is not
visible in the \ba data shown above, due to polarization effects.
However, it is the longitudinal mode that enables a {\it
two}-magnon transverse-polarized continuum excitations with a gap
$\Delta_c=2\Delta$ . Indeed, a transverse state can be constructed
from one longitudinal and one transverse magnon. In other words,
the fact that the continuum in \ba\ starts at $2\Delta$ can be
taken as an indirect evidence for the longitudinal mode. In the
future it will be very important to perform neutron experiments in
a different scattering geometry, perhaps using polarization
analysis, to observe the longitudinal mode in \ba directly, to
corroborate the remarkable results on KCuF$_3$.

In summary, the seemingly simple model of weakly interacting spin
chains demonstrates such fundamenatal phenomena of many-body
quantum mechanics as mass generation, spinon confinement, and
energy separation of ``classical'' and ``quantum''spin dynamics.
Studies of KCuF$_3$ and \ba\ shed light on the nebulous regime
where 1D quantum physics meets 3D ``classical'' magnetism and
provide the experimental basis for some very sophisticated
theoretical studies.

I would like to thank  K. Uchinokura, T. Masuda, S. Raymond,  M.
Kenzelmann,  Y. Sasago, I. Tsukada, E. Ressouche, K. Kakurai, P.
Boeni, S.-H. Lee and R. Coldea for their invaluable contributions
to the \ba\ project, S. E. Nagler and D. A. Tennant for sharing
their most recent findings on KCuF$_3$, A. Tsvelik, I. Zaliznyak,
C. L. Broholm and L.-P. Regnault for enlightening discussions and
G. Shirane for his guidance and mentorship. Work at Brookhaven
National Laboratory was carried out under Contract No.
DE-AC02-98CH10886, Division of Material Science, U.S.\ Department
of Energy.


\begin{thebibliography}{10}

\bibitem{dimers}
A. Furrer and H. U. Gudel, Phys. Rev. Lett. {\bf 39}, 657 (1977);
Y. Sasago, K.
  Uchinokura, A. Zheludev and G. Shirane, Phys. Rev. B {\bf 55}, 8357 (1997) ;
  M. Matsuda, T. Yosihama, K. Kakurai and G. Shirane, Phys. Rev. B 59, 1060
  (1999); G. Xu, C. Broholm, D. H. Reich and M. A. Adams, Phys. Rev. Lett. {\bf
  84}, 4465 (2000).

\bibitem{Haldanegap}
W. J. L Buyers, R. M. Morra, R. L. Armstrong, M. J. Hogan, P.
Gerlach, and K.
  Hirakawa, Phys. Rev. Lett. {\bf 56}, 371 (1986); R. M. Morra, W. J. L Buyers,
  R. L. Armstrong and K. Hirakawa, Phys. Rev. B {\bf 38}, 543 (1988); J. P.
  Renard, M. Verdaguer, L. P. Regnault, W. A. C. Erkelens, J. Rossat-Mignod and
  W. G. Stirling, Europhys. Lett. {\bf 3}, 949 (1987); S. Ma, C. Broholm, D. H.
  Reich, B. J. Sternlieb, and R. W. Erwin, Phys. Rev. Lett. {\bf 69}, 3571
  (1992); L.-P. Regnault, I. Zaliznyak, J. P. Renard and C. Vettier, Phys. Rev.
  B {\bf 50}, 9174 (1994).

\bibitem{continuum}
S. E. Nagler, D. A. Tennant, R. A. Cowley, T. G. Perring, and S.
K. Satija
  Phys. Rev. B 44, 12361-12368 (1991); D. A. Tennan, D. A. Tennant, T. G.
  Perring, R. A. Cowleyand S. E. Nagler, Phys. Rev. Lett. {\bf 70}, 4003
  (1993); D. A. Tennant, R. A. Cowley, S. E. Nagler and A. M. Tsvelik, Phys.
  Rev. B {\bf 52}, 13368 (1995).

\bibitem{Dender96}
D. C. Dender, D. Davidovic, D. H. Reich, C. Broholm, K. Lefmann
and G. Aeppli,
  Phys. Rev. B {\bf 53}, 2583 (1996).

\bibitem{multi-magnon}
D.A. Tennant, C. Broholm, Daniel H. Reich, S.E. Nagler,
G.E.Granroth, T.
  Barnes, G. Xu, B.C. Sales and Y. Chen, cond-mat/0005222.

\bibitem{cube}
D. C. Dender, P. R. Hammar, D. H. Reich, C. Broholm, and G.
Aeppli, Phys. Rev.
  Lett. {\bf 79}, 1750 (1997); H. M. Ronnow, M. Enderle, D. F. McMorrow, L.-P.
  Regnault, G. Dhalenne, A. Revcolevschi, A. Hoser, K. Prokes, P. Vorderwisch,
  and H. Schneider, Phys. Rev. Lett. 84, 4469 (2000).

\bibitem{cugeo3}
K. Hirota, D. E. Cox, J. E. Lorenzo, G. Shirane, J. M. Tranquada,
M. Hase, K.
  Uchinokura, H. Kojima, Y. Shibuya, and I. Tanaka, Phys. Rev. Lett. {\bf 73},
  736 (1994); O. Fujita, J. Akimitsu, M. Nishi and K. Kakurai, Phys. Rev. Lett.
  {\bf 74}, 1677 (1995); M. Aïn, J. E. Lorenzo, L. P. Regnault, G. Dhalenne, A.
  Revcolevschi, B. Hennion, and Th. Jolicoeur, Phys. Rev. Lett. {\bf 78}, 1560
  (1997) and references therein.

\bibitem{Haldanecontinuum}
I. Zaliznyak, S.-H. Lee and S. V. Petrov, Phys. Rev. Lett. {\bf
87}, 017202
  (2001).

\bibitem{Haldanefield}
Y. Chen, Z. Honda, A. Zheludev, C. Broholm, K. Katsumata, and S.
M. Shapiro,
  Phys. Rev. Lett. {\bf 86}, 1618 (2001); M. Enderle, L.-P. Regnault, C.
  Broholm, D. H. Reich, I. Zaliznyak, M. Sieling, B. L.uthi,- HMI experimental
  reports (2000); A. Zheludev, Z. Honda, K. Katsumata , R. Feyerherm and K.
  Prokes, Europhys. Lett. (in press), cond-mat/0104311.

\bibitem{Haldanestaggered}
A. Zheludev, E. Ressouche, S. Maslov, T. Yokoo, S. Raymond, J.
Akimitsu, Phys.
  Rev. Lett. {\bf 80}, 3630 (1998); S. Maslov and A. Zheludev, Phys. Rev. Lett.
  {\bf 80}, 5786 (1998); S. Raymond, T. Yokoo, A. Zheludev, S. E. Nagler, A.
  Wildes, J. Akimitsu, Phys. Rev. Lett. {\bf 82}, 2382 (1999); A. Zheludev, S.
  Maslov, T. Yokoo, J. Akimitsu, S. Raymond, S. E. Nagler, J. Phys.: Condens.
  Matter (in press), cond-mat/0006350.

\bibitem{YBANOdoping}
G. Xu, G. Aeppli, M. E. Bisher, C. Broholm, J. F. DiTusa, C. D.
Frost, T. Ito,
  K. Oka, R. L. Paul, H. Takagi, and M. M. J. Treacy, Science {\bf 289}: 419
  (2000).

\bibitem{PBNI}
Y. Uchiyama, Y. Sasago, I. Tsukada, K. Uchinokura, A. Zheludev, T.
Hayashi, N.
  Miura and P. Boni, Phys. Rev. Lett. {\bf 83}, 632-635 (1999); A. Zheludev, T.
  Masuda, K. Uchinokura and S. E. Nagler, Phys. Rev. B (in press);
  cond-mat/0103546.

\bibitem{CUGEOdoping}
Y. Sasago, N. Koide, K. Uchinokura, M. C. Martin, M. Hase, K.
Hirota, and G.
  Shirane, Phys. Rev. B 54, R6835 (1996); L.P. Regnault, J.P. Renard, G.
  Dhalenne, and A. Revcolevschi, Europhys. Lett {\bf 32} (1995) 579.

\bibitem{charge}
Y. Fujii, H. Nakao, T. Yosihama, M. Nishi, K. Nakajima, K.
Kakurai, M. Isobe,
  Y. Ueda, and H. Sawa, J. Phys. soc. Jpn. {\bf 66}, 326 (1997); B. Grenier, O.
  Cepas, L. P. Regnault, J. E. Lorenzo, T. Ziman, J. P. Boucher, A. Hiess, T.
  Chatterji, J. Jegoudez, and A. Revcolevschi , Phys. Rev. Lett. {\bf 86}, 5966
  (2001), and references therein.

\bibitem{frustrated}
See, for example, C. Broholm, G. Aeppli, G. P. Espinosa, and A. S.
Cooper,
  Phys. Rev. Lett. {\bf 65}, 3173 (1990); S.-H. Lee, C. Broholm, T. H. Kim, W.
  Ratcliff II, and S-W. Cheong, Phys. Rev. Lett. {\bf 84}, 3718 (2000), and
  references therein.

\bibitem{Bethe31}
H.~A. Bethe, Z. Phys. {\bf 71},  256  (1931).

\bibitem{Zaliznyak99}
I. A. Zaliznyak and C. Broholm and M. Kibune and M. Nohara and H.
Takagi, Phys.
  Rev. Lett. {\bf 83}, 5370 (1999).

\bibitem{DCP}
J. des Cloizeaux and J.~J. Pearson, Phys. Rev. {\bf 128}, 2131
(1962).

\bibitem{Fadeev81}
L.~D. Fadeev and L.~A. Takhtajan, Phys. Lett. {\bf 85 A},  375
(1981).

\bibitem{Muller}
G. Muller, H. Thomas, M. W. Puga, and H. Beck, J. Phys. C: Solid
State Phys.
  {\bf 14}, 3399 (1981).

\bibitem{Haldane93}
F.~D.~M. Haldane and M.~R. Zirnbauer, Phys. Rev. Lett. {\bf 71},
4055 (1993).

\bibitem{Karbach97}
M. Karbach, G. Muller, A. H. Bougourzi, A. Fledderjohann and K.-H.
Mutter,
  Phys. Rev. B {\bf 55}, 12510 (1997).

\bibitem{Scalapino75}
D. J. Scalapino, Y. Imry and P. Pincus, Phys. Rev. B {\bf 11},
2042 (1975).

\bibitem{Schulz96}
H.~J. Schulz, Phys. Rev. Lett. {\bf 77}, 2790 (1996).

\bibitem{Essler97}
F.~H.~L. Essler, A.~M. Tsvelik, and G. Delfino, Phys. Rev. B {\bf
56}, 11001
  (1997).

\bibitem{Tennant95}
D.~A. Tennant, S. E. Nagler, D Welz, G. Shirane and K. Yamada,
Phys. Rev. B
  {\bf 52}, 13381 (1995).

\bibitem{Tsukada99}
I. Tsukada, Y. Sasago, K. Uchinokura, A. Zheludev, S. Maslov, G.
Shirane, K.
  Kakurai and E. Ressouche, Phys. Rev. B {\bf 60}, 6601 (1999).

\bibitem{Kenzelmann01}
M. Kenzelmann, A. Zheludev, S. Raymond, E. Ressouche, T. Masuda,
P. Böni, K.
  Kakurai, I. Tsukada, K. Uchinokura and R. Coldea, Phys. Rev. B (in press),
  cond-mat/0012452.

\bibitem{Zheludev00Ba}
A. Zheludev, M. Kenzelmann, S. Raymond, E. Ressouche, T. Masuda,
K. Kakurai, S.
  Maslov, I. Tsukada, K. Uchinokura and A. Wildes, Phys. Rev. Lett. {\bf 85},
  4799 (2000).

\bibitem{Zheludev01Ba}
A. Zheludev and M. Kenzelmann and S. Raymond and T. Masuda and K.
Uchinokura
  and S.-H. Lee, cond-mat/0105223.

\bibitem{Lake00}
B. Lake, D.~A. Tennant, and S.~E. Nagler, Phys. Rev. Lett. {\bf
85},  832
  (2000).

\bibitem{Nagler01}
S. E. Nagler and D. A. Tennant, private communication.

\end{thebibliography}

\end{document}